\documentclass[prl,twocolumn,amsmath,amssymb,showpacs,floatfix,nofootinbib,preprintnumbers,superscriptaddress]{revtex4-1}
\usepackage[latin1]{inputenc}
\usepackage{amsmath}
\usepackage{amsfonts}
\usepackage{amssymb}
\usepackage{graphicx}
\usepackage{dcolumn}
\usepackage{subcaption}
\usepackage{caption}
\usepackage{float}
\usepackage{bm}
\usepackage{latexsym}
\usepackage{epsfig}
\usepackage[normalem]{ulem}
\usepackage{graphicx}
\usepackage[colorlinks=true]{hyperref}

\begin{document}
\title{Multi-messenger tests of gravity with weakly lensed gravitational waves}
\author{Suvodip Mukherjee}\email{mukherje@iap.fr}
\affiliation{Center for Computational Astrophysics, Flatiron Institute, 162 5th Avenue, 10010, New York, NY, USA}
\affiliation{Institut d'Astrophysique de Paris (IAP), UMR 7095, CNRS/UPMC Universit\'e Paris 6, Sorbonne Universit\'es, 98 bis boulevard Arago, F-75014 Paris, France}
\affiliation{ Institut Lagrange de Paris (ILP), Sorbonne Universit\'es, 98 bis Boulevard Arago, 75014 Paris, France}
\author{Benjamin D. Wandelt}\email{bwandelt@iap.fr}
\affiliation{Center for Computational Astrophysics, Flatiron Institute, 162 5th Avenue, 10010, New York, NY, USA}
\affiliation{Institut d'Astrophysique de Paris (IAP), UMR 7095, CNRS/UPMC Universit\'e Paris 6, Sorbonne Universit\'es, 98 bis boulevard Arago, F-75014 Paris, France}
\affiliation{ Institut Lagrange de Paris (ILP), Sorbonne Universit\'es, 98 bis Boulevard Arago, 75014 Paris, France}
\affiliation{Department of Astrophysical Sciences, Princeton University, Princeton, NJ, 08540, USA}
\author{Joseph Silk}\email{joseph.silk@physics.ox.ac.uk}
\affiliation{Institut d'Astrophysique de Paris (IAP), UMR 7095, CNRS/UPMC Universit\'e Paris 6, Sorbonne Universit\'es, 98 bis boulevard Arago, F-75014 Paris, France}
\affiliation{ Institut Lagrange de Paris (ILP), Sorbonne Universit\'es, 98 bis Boulevard Arago, 75014 Paris, France}
\affiliation{The Johns Hopkins University, Department of Physics \& Astronomy, \\ Bloomberg Center for Physics and Astronomy, Room 366, 3400 N. Charles Street, Baltimore, MD 21218, USA}
\affiliation{Beecroft Institute for Cosmology and Particle Astrophysics, University of Oxford, Keble Road, Oxford OX1 3RH, UK
}
\date{\today}
\begin{abstract}
General relativity (GR) predicts concordant trajectories for photons and gravitational waves (GW). We propose a new multi-messenger avenue (GW-CMB-CMB) to prove this aspect of fundamental physics by cross-correlating the GW signal of astrophysical origin with the lensing field derived from the cosmic microwave background (CMB).  This new window will allow robust measurement of the prediction from GR with high signal-to-noise and will be able to unveil the true nature of gravity using the GW sources detected by missions such as the  Laser Interferometer Space Antenna (LISA), Einstein Telescope and Cosmic Explorer.
\end{abstract}
\pacs{}
\maketitle
\paragraph{Introduction :}
Our endeavour to understand the Universe through Electromagnetic Waves (EWs) 
over a wide frequency band ranging from radio to gamma-rays has enabled us to construct the standard model of cosmology over a large redshift range and unveiled various cosmic secrets. Several probes such as supernovae, the Cosmic Microwave Background (CMB), galaxy surveys, and quasars are the founding pillars of the  Lambda Cold Dark Matter (LCDM) standard model of the Universe. This model matches the observational evidence of enigmatic late-time acceleration and invisible, dark matter. 

Gravitational waves (GWs)
 are a new avenue capable of probing the Universe through white dwarfs,  neutron stars,  {and} binary black holes (BBHs).  {In the framework of General Relativity}, GWs (like EWs) propagate along the geodesics defined by the Friedmann-Lemaitre-Robertson-Walker (FLRW) metric. However, due to the presence of the matter distribution in the Universe, GWs interact gravitationally with the matter distribution and hence propagate through the perturbed FLRW metric, written as
\begin{equation}\label{flrw-1}
ds^2= (1-2\Phi)dt^2 - a(t)^2(1-2\Psi)(dx^2 + dy^2 +dz^2),
\end{equation}
where, $a(t)$ is the scale factor, $\Phi$ \& $\Psi$ are the scalar perturbations due to the matter distribution in the Universe. 

\begin{figure}
    \centering
        \includegraphics[width=1.\linewidth,keepaspectratio=true]{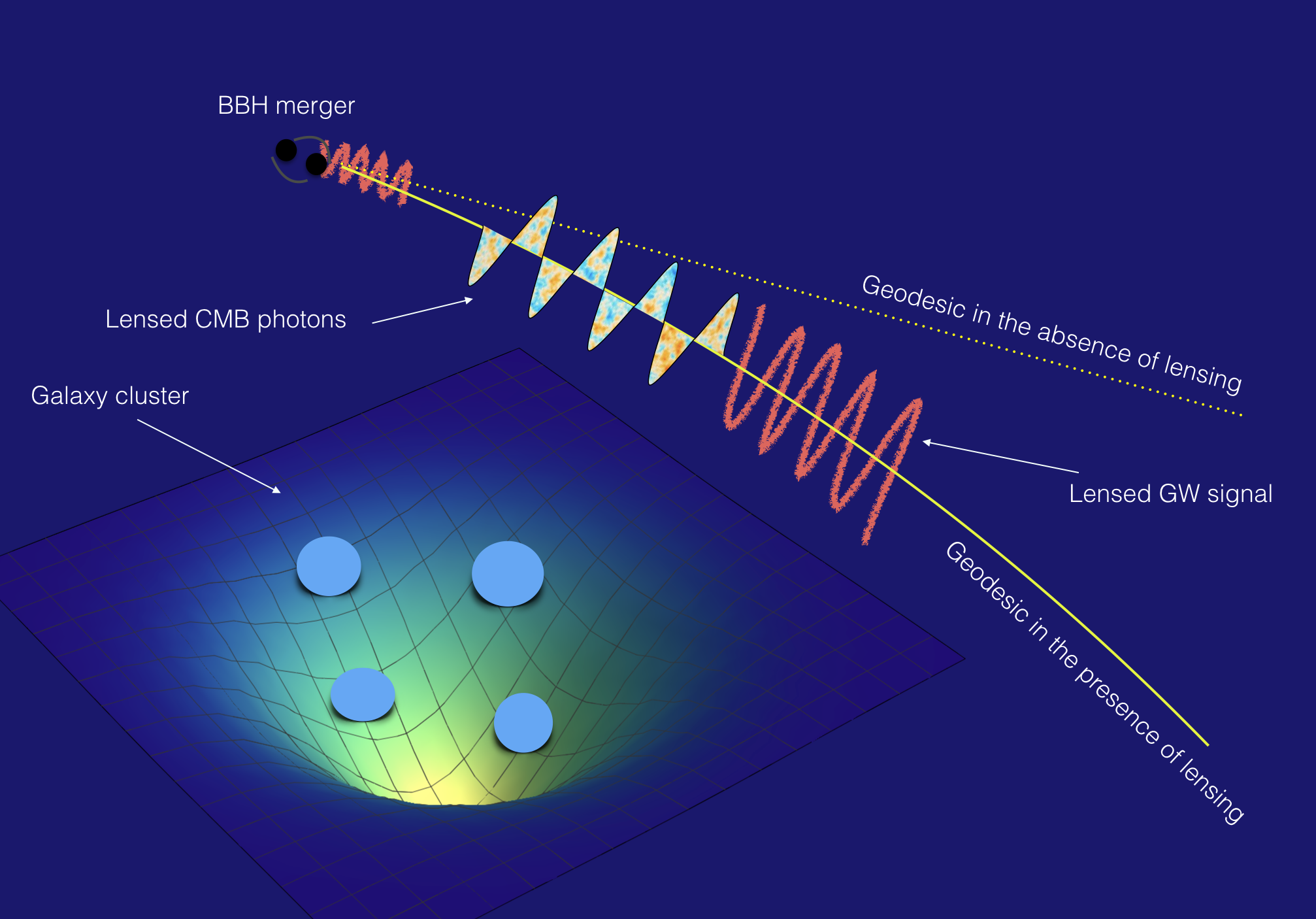}
             \captionsetup{singlelinecheck=on,justification=raggedright}
           \caption{{GR predicts identical geodesics of the lensed GW as for the lensed photons due to cosmic perturbations and hence should be correlated. This schematic diagram depicts the physical mechanism behind the correlation between the lensed CMB photons and the lensed GW signal from astrophysical sources.} \label{fig-cmbgwsc}}
 \end{figure}
 
We propose a new method for {probing} 
 {the propagation of GWs in a perturbed metric. General relativity predicts that weak lensing due to the intervening matter distribution affects the geodesics of EWs and GWs in the same manner\cite{1987thyg.book.....H} (see the schematic diagram in Fig.~\ref{fig-cmbgwsc}). To test this fundamental prediction, the cross-correlation of GW and EW signals we propose is the \textit{only known avenue} capable to detect the \textit{gravitational lensing of GW} unambiguously.}

 {Furthermore, this new avenue will be capable to explore multiple theoretical aspects such as (i) alternative theories to GR \cite{Saltas:2014dha,Nishizawa:2017nef} (by probing the running of the Planck mass, anisotropic stress, graviton mass); (ii) dimensions of the space-time \cite{Cardoso:2002pa}; and (iii) the difference between the two scalar potentials $\Phi$ and $\Psi$ \cite{Carroll:2006jn, Bean:2006up, Hu:2007pj, Schmidt:2008hc, Silvestri:2013ne}.  {The above mentioned theoretical aspects lead to several observable effects. Alternative theories of gravity lead to a different propagation equation of GWs from GR which can be written in terms of a few additional parameters $\alpha_M, c_T, \mu,$ and $\Gamma_{ij}$ \cite{Lombriser:2015sxa, Lombriser:2016yzn, Sakstein:2017xjx, PhysRevLett.119.251301, Nishizawa:2017nef, Belgacem:2017ihm} as }
 \begin{equation}\label{gwprop}
 h_{ij}'' + (2 + \alpha_M)\mathcal{H}h_{ij}' + (c_T^2k^2 + a^2\mu^2)h_{ij}= a^2\Gamma_{ij}.
\end{equation}
 {The single combined
EW/GW observation of a binary neutron star ruled out vast swathes of alternative gravity models via the  time-delay between EW and GW signals \cite{PhysRevLett.119.251301} and constraining the term $c_T$ and $\mu$ which are related to the speed of GW and the graviton mass. However, the other two parameters such as $\alpha_M$ and $\Gamma_{ij}$ remain unconstrained. Along with the effects on GW propagation, alternate theories of gravity also affect the Poisson equation} 
 \begin{align}\label{pois}
 \begin{split}
\nabla^2 (\Phi+ \Psi)=& 8\pi G_{light}a^2\rho \delta,\\
\nabla^2 \Phi=& 4\pi G_{matter}a^2\rho \delta,
\end{split}
\end{align}
 {where $G_{light}$ and $G_{matter}$ are equal in GR, but differ in alternative theories of gravity, $\rho$ is the matter density and $\delta$ denotes the density contrast. The gravitational lensing of GW is affected by both $\Phi$ and $\Psi$. As a consequence, in order to understand the theory of gravity it is essential to measure all the effects on the GW signal as it propagates through the space-time. For a given theory of gravity, all the parameters $\alpha_M$,  $\Gamma_{ij}$, $G_{matter}$ and $G_{light}$ lead to an observable signature and are not necessary to be independent from each other.  The method proposed by us in this paper enables to probe both Eq. \eqref{gwprop} and \eqref{pois}. As a result, the modification in the luminosity distance due to the modified theories of gravity and the effects on the GW strain (which can be related to the luminosity distance, see the discussion around Eq. \eqref{gw-del-dl}) can be jointly estimated in this method. It will also be a direct test of the equivalence principle by comparing the concurrent trajectory of EW and GW up to a high cosmological redshift.} {This method opens a new scientific window to study fundamental physics with  GW binaries by exploring the two-point correlation between GW strain and CMB lensing (or equivalently the three-point correlation between GW strain and CMB fields such as temperature and polarization anisotropies) and goes beyond the luminosity distance-redshift test which probes only the background cosmology. It also enables a correct estimate of the signatures of modified gravity theories from the strain of the GW signal by eliminating the degeneracy with the weak lensing field. Finally, our proposed scheme makes it possible to remove the effect of weak lensing (or \textit{delense}) from  the GW signal and hence reduce the additional uncertainty in the GW source parameters.}
  
\paragraph{Weak lensing of CMB :}\label{cmb-lensing}
Cosmological observables like the CMB  and the galaxy field carry the imprint of lensing and were measured recently \cite{Ade:2015zua}.  
The weak lensing of the CMB temperature ($T$) and the polarization field ($E,\, B$)  can be written as 
\begin{align}\label{cmb-lens1}
\begin{split}
\tilde X(\hat n)= X(\hat n +  \vec \bigtriangledown \phi (\hat n)),\\
\end{split}
\end{align}
where $X \in {T, E, B}$ and $\vec \bigtriangledown\phi (\hat n)$ is the deflection angle and $\phi (\hat n)$ is the lensing potential. 
Different lensing estimators \cite{Okamoto:2003zw, Carron:2017mqf, Millea:2017fyd} are developed to reconstruct this signal by using the off-diagonal correlations between $T,\,E,\,B$ ($EB, TT, EE, TE, BB, TB$). The commonly used quadratic minimum variance estimator \cite{Okamoto:2003zw} reconstructs the lensing potential with the corresponding lensing reconstruction noise $N_l^{\kappa\kappa}$ \cite{Okamoto:2003zw} which is obtained from all the cross \& autocorrelations between $T,\, E,\, B$ mentioned above. 

The deflection angle $\vec \bigtriangledown \phi$ is related to the lensing convergence field $\kappa$ by the relation $\kappa = -\bigtriangledown^2 \phi/2$ (or in the spherical harmonic space \footnote{Any spin$-0$ field $P(\hat n)$ in the sky can be written in the spherical harmonics basis as $P(\hat n) = \sum_{lm} P_{lm} Y_{lm}(\hat n)$.} by the relation $\kappa_{lm} = l(l+1)\phi_{lm}/2$). This in turn is related to the intervening matter distribution between us and the CMB source plane by the relation
 \begin{align}\label{cmb-kappa}
\begin{split}
\kappa_{CMB} (\hat n) =& \int_{0}^{z_s} d z\, W_{cmb}(\chi(z)) \delta(\chi(z)\hat n, z),
\end{split}
\end{align}
where $z_s=1089$ is the source redshift of CMB, and $W_{cmb} (\chi (z))$ is defined as
\begin{align}\label{cmb-kappa-a}
\begin{split}
W_{cmb} (\chi (z)) =  \frac{3}{2}\frac{\Omega_m H_0^2(1+z)\chi(z)}{cH(z)}\bigg[ \frac{(\chi(z_s)-\chi(z))}{(\chi(z_s))}\bigg],\end{split}
\end{align}
where, $H_0$ is the Hubble constant, $\Omega_m$ is the matter density, $\chi(z)$ is the comoving distance to  redshift $z$ and $\delta$ is the perturbation in the matter distribution. Here we have used the Poisson equation, $\bigtriangledown^2\Phi = 3 \Omega_{m0}H_0^2\delta/2a$ to connect the potential $\Phi$ with the matter density perturbations $\delta$. 
\paragraph{Effect of cosmological perturbations on GWs :}
GW strain from the inspiraling binaries can be written in Newtonian order \footnote{For this paper, we will use the Newtonian waveform \cite{Cutler:1994ys}, which can be generalized to  the waveform including Post-Newtonian (PN) corrections and also with the waveform generated using numerical relativity. Inclusion of the PN corrections will not change the primary concept of this paper.}  
in frequency domain ($\nu$) as \cite{Poisson:1995ef, Cutler:1994ys, 1987thyg.book.....H, maggiore2008gravitational}
\begin{align}\label{gw-waveform}
\begin{split}
h (\nu_z)&= \mathcal{Q}(\text{angles})\sqrt{\frac{5}{24}}\frac{G^{5/6}M_z^2 (\nu_zM_z)^{-7/6}}{c^{3/2}\pi^{2/3}d_L} e^{i\phi_z},
\end{split}
\end{align}
where $\nu_z$ and $M_z$ are redshifted frequency and redshifted chirp mass respectively which are related to the source chirp mass $M_c$ and emitted frequency $\nu_e$ by $M_z= (1+z) M_c$ and $\nu_z=\nu_e/(1+z)$.  $\phi_z=  2\pi\nu_zt_0 + \phi_r(t_0)$ is the phase of the GWs with $t_0$ as the stationary point of the phase. $\mathcal{Q}(\text{angles})$ is the factor which depends upon the source orientation. The above equation is valid only in the  inspiraling  phase of the binaries and not during its merger. The information regarding the background cosmology can be inferred from the luminosity distance independently of the chirp mass by using the relationship \cite{1986Natur.323..310S}
\begin{equation}\label{gw-dl}
\begin{split}
d_L \propto & \frac{1}{\bar h(t) \tau \nu^2},\, 
\text{where } \tau \equiv  \bigg(\frac{d\nu/dt}{\nu}\bigg)^{-1} \propto \frac{\pi M_z^2}{(\pi M_z)^{11/3}\nu^{8/3}},\\& \text{and } \bar h(t) \propto\frac{M_z (\pi \nu_zM_z)^{2/3}}{d_L}.
\end{split}
\end{equation}
Here $\bar h$ is the GW strain averaged over detectors and source orientations and  $\tau$ is the time-scale related to the change of the frequency. The above quantity is  independent of the chirp mass and is an useful estimator of the luminosity distance \cite{1986Natur.323..310S}. A more general luminosity distance estimator using the two polarization states of the GW ($h_+, h_\times$) and the source orientation has also been studied \cite{2010ApJ...725..496N}.

The observed GW strain can be modeled as $h^{\text{obs}}(t)= h (t) + n(t) $ 
 where $h (t)$ is the signal strain and $n(t)$ is the noise strain. The signal-to-noise ratio ($\rho$) for the GW signal can be written in the frequency domain as \cite{Flanagan:1997sx, maggiore2008gravitational} \footnote{The factor 4 appears due to the definition of the SNR with only one-sided noise density. Usually the characteristic strain $h_c (\nu) \equiv 2\nu h(\nu)$ is used in the literature. }
\begin{align}\label{gw-signal-2}
\begin{split}
\rho^2= 4 \int_0^\infty \frac{\nu^2|h(\nu)|^2}{|h_n(\nu)|^2} d\ln\nu,
\end{split}
\end{align}
where  $h_n$ is the dimensionless noise strain which depends upon the experimental noise power spectrum ($S_n$) as $|h_n(\nu)|^2= \nu S_n$. $|h(\nu)|^2$ is the power spectrum of the signal strain and depends upon the GW source properties and luminosity distance as mentioned in Eq. \eqref{gw-waveform}. 

GWs  propagate through the geodesics defined by the perturbed FLRW metric defined in Eq. \eqref{flrw-1}. The presence of cosmological perturbations in this metric leads to change in the emitted GW frequency $\tilde \nu$ which can be written as \cite{Laguna:2009re}
 \begin{equation}\label{gw-freq}
\tilde \nu= \nu \bigg(1- \bigg(\Phi |_{e}^{r} - (\vec n.\vec v)|_{e}^{r} - \int_{\lambda_{e}} ^{\lambda_{r}} \partial_\eta (\Psi +\Phi) d\lambda'\bigg)\bigg), 
\end{equation}
where, the first term is the Sachs-Wolfe (SW) effect,  the second term is the Doppler effect due to the difference in the velocity of the source and the observer, and the third term is the Integrated-SW (ISW) effect due to the presence of dark energy.  The RMS fluctuations in $\nu$ is of the order $10^{-5}$ \cite{Laguna:2009re} and can be considered to be a negligible effect for the current discussion.

The GW strain also gets modified by the matter perturbations \cite{Laguna:2009re, Camera:2013xfa, Takahashi:2005ug,Bertacca:2017vod}, with the dominant contributions arising from  lensing for the GW sources at high redshift 
\begin{align}\label{gw-amp-a}
\begin{split}
\tilde h(\hat n, \nu_z)=h(\nu_z)[1 + \kappa_{gw} (\hat n)],
\end{split}
\end{align}
where $\kappa_{gw} (\hat n)$ is the convergence field due to weak lensing, which can be written in terms of the  {intervening matter density field $\delta$ by the relation}
\begin{align}\label{gw-kappa-2}
\begin{split}
\kappa_{gw} (\hat n) = & \int_{0}^{z_s} d z\, W_{gw}(\chi(z)) \delta(\chi(z)\hat n, z),
\end{split}
\end{align}
where $W_{gw}(\chi(z))$ is the lensing kernel defined as
\begin{align}\label{gw-kappa-2a}
\begin{split}
W_{gw} (\chi(z)) =& \frac{3}{2}\frac{\Omega_m H_0^2(1+z)\chi(z)}{cH(z)} \\& \times \int_z^\infty dz' \frac{dn_{gw}(z')}{dz'}\frac{(\chi(z')-\chi(z))}{(\chi(z'))}.
\end{split}
\end{align}
Here $\frac{dn_{gw}(z)}{dz}$ is the normalised ($\int dz \frac{dn_{gw}(z)}{dz} =1$) redshift distribution of the GW sources, $\chi(z)$ is the comoving distance to redshift $z$, $\Omega_m\equiv \rho_m/\rho_c$ is the matter density in terms of critical density $\rho_c$ and $H_0$ is the current value of the Hubble parameter. 

The all-sky lensing convergence map can be expressed as $\kappa(\hat n)= \sum_{lm} \kappa_{lm} Y_{lm}(\hat n)$, where $Y_{lm}(\hat n)$ are the spherical harmonics basis.
The correlation between the convergence lensing field of the CMB and GWs due to the perturbed geodesics can be written as 
\begin{align}\label{gw-cmb-corr-2}
\begin{split}
C_l^{\kappa_{gw} \kappa_{cmb}} \equiv  \langle  (\kappa_{{gw}})_{lm} (\kappa^*_{{cmb}})_{l'm'} \rangle \delta_{ll'}\delta_{mm'},
\end{split}
\end{align}
where the angular bracket denotes ensemble average and $ C^{\kappa_{gw}\kappa_{cmb}}_l$ is the power spectrum of the cross-correlation field between CMB and GWs which can be written as
\begin{align}\label{gw-cmb-corr-2a}
\begin{split}
C^{\kappa_{gw}\kappa_{cmb}}_l =&\int \frac{d z}{\chi(z)^2} \frac{H(z)}{c} \bigg[W_{gw}(\chi(z))W_{cmb}(\chi(z)) \\& \times P_{\delta}((l+1/2)/\chi(z))\bigg].
\end{split}
\end{align}
Here $P_{\delta}((l+1/2)/\chi)$ is the non-linear matter power spectrum of the cosmic density field evaluated at $k=(l+1/2)/\chi (z)$ in the Limber approximation, obtained using the numerical code CLASS \cite{2011arXiv1104.2932L, 2011JCAP...10..037A, 2011JCAP...07..034B}.  This correlation is observable as a GW-CMB-CMB three-point correlation since the estimate of the CMB lensing potential is quadratic in the CMB anisotropies. The predicted root mean square (RMS) signal strength $\Delta_{GW-CMB}\equiv [\sum _l (2l+1)C^{\kappa_{gw}\kappa_{cmb}}_l /4\pi]^{1/2}$ for the LCDM model in the framework of general relativity is shown as a function of redshift $z$ of the source in Fig. \ref{fig-cmbgw}.  We have taken the GW source redshift distribution as $\frac{dn}{dz}= \delta (z-z')$  for the plot in Fig. \ref{fig-cmbgw}. The theoretical signal strength of CMB lensing-GW correlation for LCDM model is greater than $10^{-2}$ at redshift above $0.5$.  
Since the CMB source redshift is at $z_s=1089$, the GW sources present at high redshift have more overlap with the CMB lensing kernel and hence exhibit a stronger signal. The auto-correlation between the GW signal is also depicted in Fig. \ref{fig-cmbgw}, in accordance with \cite{Laguna:2009re}. Along with CMB lensing-GW correlation, galaxy-GW cross-correlation is  another avenue to study the lensing of GW strain \cite{Mukherjee:2019galgw}. 
\begin{figure}
\centering
 \includegraphics[width=1.\linewidth,keepaspectratio=true]{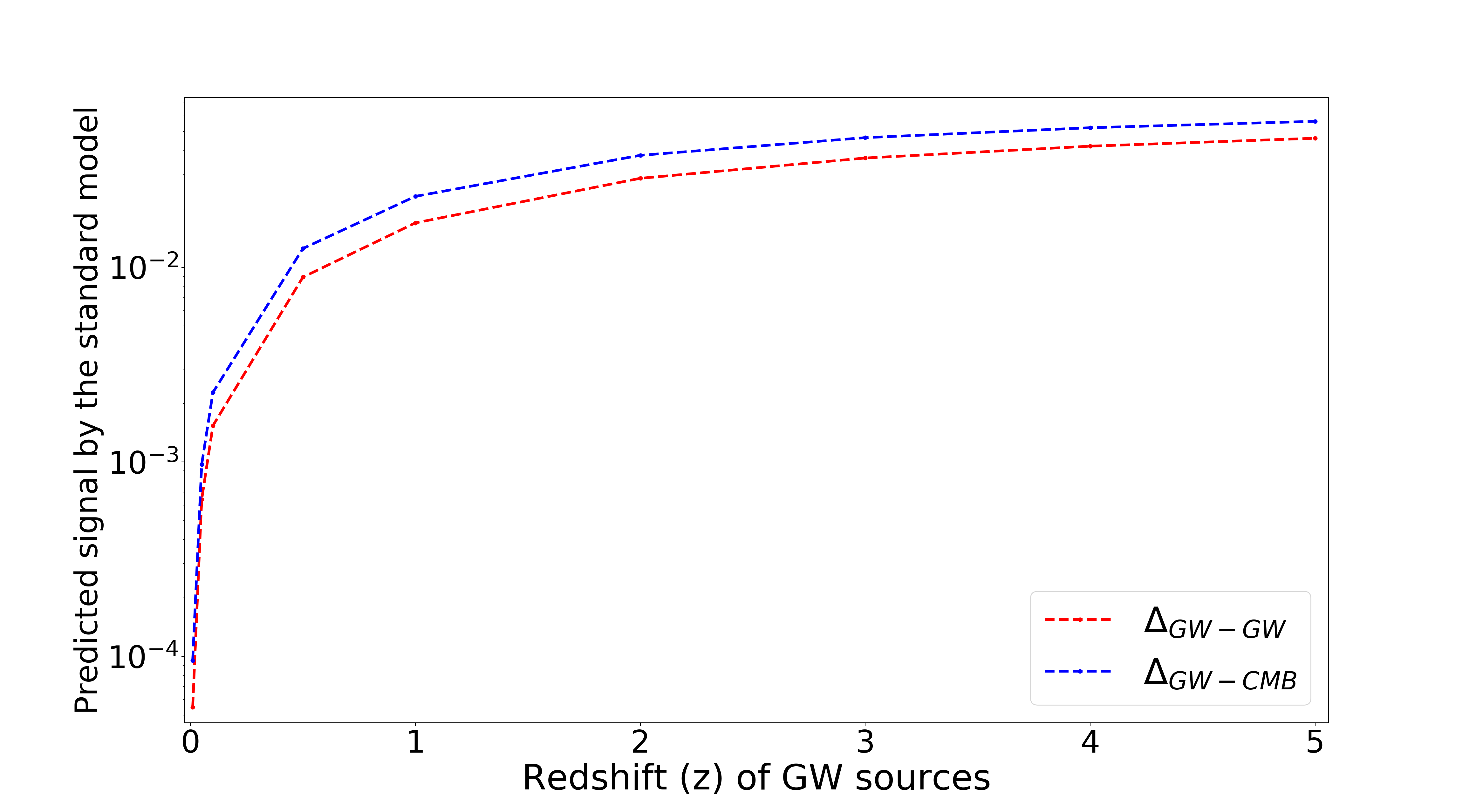}
 \captionsetup{singlelinecheck=on,justification=raggedright}
\caption{The theoretical RMS  signal of the CMB lensing-GW correlation and GW-GW correlation due to the convergence field is shown in blue and red respectively as a function of redshift for best-fit cosmological parameters using the non-linear matter power spectrum.\label{fig-cmbgw}}
\end{figure}

\paragraph{Estimator of the convergence field from GW strain :}  {In order to make an estimate of the convergence field from the luminosity distance, we need to make an
 estimate of the true luminosity distance $d^{es}_L$ to the GW source by using the redshift $z$ of the source and the best-fit cosmological parameters. 
 The massive BBHs which can be detected by  LISA are expected to have  electromagnetic signatures \cite{2041-8205-752-1-L15,Haiman:2018brf,Palenzuela:2010nf,Farris:2014zjo,Gold:2014dta,Armitage:2002uu} and we can identify the host galaxy \cite{Petiteau:2011we} and its redshift ($z$) using upcoming missions \cite{2009arXiv0912.0201L, Padovani:2017epc, ELT, Maartens:2015mra} or other dedicated spectroscopic surveys. For stellar origin BBHs which can be probed from Cosmic Explorer, we may not have an electromagnetic counter-part and as a result, the redshift error will be large.
 
 So, by using the redshift from the electromagnetic follow-up and the value of the background cosmological parameters from large scale structure and CMB upcoming missions \cite{2010arXiv1001.0061R, 2009arXiv0912.0201L, Dore:2018smn, Abazajian:2016yjj}, \footnote{These cosmological probes will reach a much better accuracy in the cosmological parameters than the current estimates.}}
 we can estimate the luminosity distance $d^{\text{es}}_L$  as  
\begin{equation}
d^{\text{es}}_L= \frac{c}{H_0}(1+\bar z)\int_0^{\bar z}\frac{dz'}{\sqrt{\bar  \Omega_m(1+z')^3 + \bar\Omega_{de}}}\label{dl-es}.
\end{equation}
By using Eq. \eqref{gw-dl} and \eqref{gw-amp-a}, we can write the estimator of the convergence field for a distribution of the GW sources as 
\begin{align}\label{gw-del-dl}
\begin{split}
\mathcal{\hat D}_L (\hat n)&= \kappa_{gw} (\hat n) + \epsilon(\hat n)  - \epsilon(\hat n) \kappa_{gw}(\hat n),
\end{split}
\end{align}
here $\mathcal{\hat D}_L(\hat n)\equiv 1-\frac{D_L (\hat n)}{d^{\text{es}}_L(\hat n)}$ is a probe to the convergence field along with an additional term which is related to the error $\epsilon= 1-d_L/d^{\text{es}}_L$) in the estimate of the true luminosity distance. We expect the error $\epsilon$ to be small ($\epsilon <<1$) if the source redshift and cosmological parameters are measured accurately. More details about the estimator can be found in \cite{Mukherjee:2019galgw}.  {Here we have assumed that the waveform of the GW signal can be modeled according to GR. Alternate theories of gravity can have imprints on the GW signal in the strong gravity regime. But it is often assumed that such effects can be avoided by a screening mechanism \cite{LIGOScientific:2019fpa}. The test proposed in this paper can be also done by including the effects in the waveform along with the effects from GW propagation and lensing. This we will address in  future work.}

{ {Studies from simulations have shown that BBH sources detected by LISA can have an electromagnetic counterpart \cite{2041-8205-752-1-L15,Haiman:2018brf,Palenzuela:2010nf,Farris:2014zjo,Gold:2014dta,Armitage:2002uu} which will allow us to obtain the redshift of these objects.} \footnote{ {However there can be scenarios where the electromagnetic counterpart is not present for the LISA source. In this case, we need to consider large redshift errors as considered for the forecast for Cosmic Explorer. }} As a result, the second term in Eq. \eqref{gw-del-dl} can be negligible for $10^4-10^7$ $M_\odot$. The error in the redshift measurement from a follow-up photometric survey is considered as $\sigma_z/(1+z) = 0.03$ in this analysis. However, the redshift error can be negligible ($\sigma_z \approx 0$) for a spectroscopic follow-up mission, resulting in an improvement in the SNR.} 
For stellar origin BBHs which can be probed from Cosmic Explorer, we may not have any  electromagnetic counter-part and as a result, the $d^{es}_l$ will be noise-dominant leading to a large value of $\epsilon$. In our analysis, we have taken this into account  by considering $100\%$ error in the estimate of source redshift for the forecast of Cosmic Explorer.

The cross-correlation $\mathcal{\hat E}^{\kappa_{gw}\kappa_{cmb}}$ between the convergence field from CMB $
\hat \kappa_{\text{CMB}} (\hat n)$ and the $\mathcal{\hat D}_L (\hat n)$  signal can be written as
\begin{align}\label{gw-cmb-corr-1}
\begin{split}
\mathcal{\hat E}^{\kappa_{gw}\kappa_{cmb}} =\int \frac{d^2\hat n}{4\pi}\, \big(\epsilon(\hat n) +\, \hat \kappa_{gw} (\hat n)\big)\hat \kappa_{\text{CMB}}(\hat n').
\end{split}
\end{align}
 The correlation of the GW strain with the convergence map is a three-point correlation function (also called the bispectrum \cite{2011JCAP...03..018L,Mangilli:2013sxa}) between two CMB fields ($T\, E,\,B$) and the GW strain.} This equation implies that the sources of GW compact objects detected at a direction $\hat n$ with a strain $\tilde h(\nu, \hat n)$ will show a correlated signal with the convergence map obtained from the CMB.  As the convergence field is uncorrelated with the error $\epsilon(\hat n)$, the first term on the right hand side goes to zero. The second term is a cosmological signal which captures the RMS fluctuations due to the convergence field of CMB and GW.
 
 The corresponding covariance matrix with a diagonal approximation can be calculated using 
\begin{align}\label{gw-cmb-noise-1}
\begin{split}
(\sigma^{gw-cmb}_l)^2 & = \frac{1}{f_{\text{sky}}(2l+1)} \bigg((C^{\kappa_\text{gw}\kappa_\text{gw}}_l + N_l^{\mathcal{DD}})(C^{\mathcal{\kappa_{\text{cmb}}}}_l \\& + N^{\kappa\kappa}_l) + (C^{\kappa_{gw}\kappa_{CMB}}_l)^2\bigg),
\end{split}
\end{align}
where, $C^{\kappa_\text{gw}\kappa_\text{gw}}_l = \langle  (\kappa_{\text{gw}})_{lm} (\kappa_{\text{gw}})_{l'm'} \rangle \delta_{ll'}\delta_{mm'}$ is the convergence power-spectrum from the auto-correlation of GW-GW, $N_l^{\mathcal{DD}}$ is the measurement error associated with the  GW luminosity distance determination (we explain this quantity in detail later),  $N_l^{\kappa \kappa}$ is the reconstruction noise due to lensing estimation \cite{Okamoto:2003zw} (as we mentioned previously) and $f_{sky}$ is the sky fraction available common between the GW sources and CMB.  
For a number of GW sources $N_{gw}$ with the same value of $\sigma_{d_l}$, the spatial GW noise mentioned in Eq. \eqref{gw-cmb-noise-1} can be written as
\begin{align}\label{gw-nldd-1}
\begin{split}
N_l^{\mathcal{DD}} = \frac{4\pi}{N_{gw}}\bigg(\frac{\sigma^2_{d_l}}{d^2_l} + \frac{\sigma^2_{b}}{d^2_l}\bigg)e^{l^2\theta_{min}^2/8\ln 2},
\end{split}
\end{align}
where $\sigma^2_{d_l}$ can be obtained using Eq. \eqref{gw-signal-2} and $\sigma^2_{b}$ is the error due to the estimation of the background cosmological parameters, lensing and redshift of the GW source. The presence of the sky-localization error for the GW sources will lead to a poor angular resolution of the GW sources, and as a result the spatial correlations cannot be probed for scales smaller than the angular scale of the sky localization error. This translates to a maximum value of $l_{max}\approx 180^\circ/\theta_{min}$, beyond which there is no signal and the CMB lensing-GW correlation is noise-dominated. 

The SNR of the CMB lensing-GW correlation  can be written in terms of $C^{\kappa_{gw}\kappa_{cmb}}_l$ and $\sigma^{gw-cmb}_l$ are defined in Eq. \eqref{gw-cmb-corr-2a} and Eq. \eqref{gw-cmb-noise-1} respectively as $(SNR)^2 = \sum^{l_{\text{max}}}_l \bigg(\frac{C^{\kappa_{gw}\kappa_{cmb}}_l}{\sigma^{gw-cmb}_l}\bigg)^2$. 
Using the LISA noise and the Newtonian GW waveform of the form mentioned in Eq. \eqref{gw-waveform}, we make a Fisher estimate of the luminosity distance error for different masses of BBHs up to a maximum frequency equal to the merger frequency defined as $\nu_{\text{merge}} = 205 (20M_\odot/M)\, \text{Hz}$ \cite{Flanagan:1997sx}. We do not consider the merger and ringdown phase of the BBHs in the estimate of the SNR for this paper. Inclusion of the merger and ring-down phase of BBHs will improve the SNR \cite{Holz:2005df}.

\paragraph{Forecast for LISA and Cosmic Explorer :}
The  {CMB lensing-GW} correlation is strong for GW sources at high redshift (as depicted in Fig. \ref{fig-cmbgw}) and hence we expect that the signal will be more easily accessible from LISA \cite{2017arXiv170200786A} and Cosmic Explorer \cite{Evans:2016mbw} than from LIGO \cite{ligo}  {(for the currently predicted merger rates \cite{Abbott:2016nhf, LIGOScientific:2018jsj}).}  {However, if the number of GW sources are more by about an order of magnitude than the current estimate of the event rate \cite{Abbott:2016nhf} then this signal is also accessible from advance-LIGO.  
In this analysis, we treat the number of GW sources $N_{\text{gw}}$ and the smallest angular scale $\theta_{min}$ as free parameters.  {The numbers of GW sources per unit redshift for four years of LISA operational time are motivated from a few recent theoretical studies \cite{2007MNRAS.380.1533M, PhysRevD.93.024003}. For the Cosmic Explorer-like survey, we have taken GW detection rate $\Delta n_{gw}/\Delta T= 24-112$ Gpc$^{-3}$ yr$^{-1}$ (comoving volume) \cite{LIGOScientific:2018jsj}}. For LISA  and Cosmic Explorer we plot the} cumulative $SNR/\sqrt{f_{sky}}$ in Fig. \ref{Fig-cmb-gw-snr} for a total mass of $2 \times 10^5$ $M_\odot$ and $50\,M_\odot$ $(M_{total}\equiv  M_1+ M_2)$ of the BBHs (with $M_{ratio} \equiv M_1/M_2 =1$) as a function of the maximum redshift (considered in the cumulative SNR), $N_{gw}$ and $\theta_{min}$. The region in cyan indicate $\text{SNR} \leq 3$ for $f_{sky} = 1$. This plot indicate that there is a large measurable window for different possibilities  of $N_{gw}, M_{BH}, \theta_{min}$ which can probe the CMB lensing-GW correlation signal with high statistical significance.  The detection of the GW sources with electromagnetic counterparts are going to improve the sky localization of the source \cite{Nissanke:2011ax}. This will result into further improvement of the lensing signal from gravitational waves. 
\begin{figure}
          \begin{subfigure}{1.\linewidth}
     \includegraphics[trim={0 0 0 0cm}, clip, width=1.\linewidth]{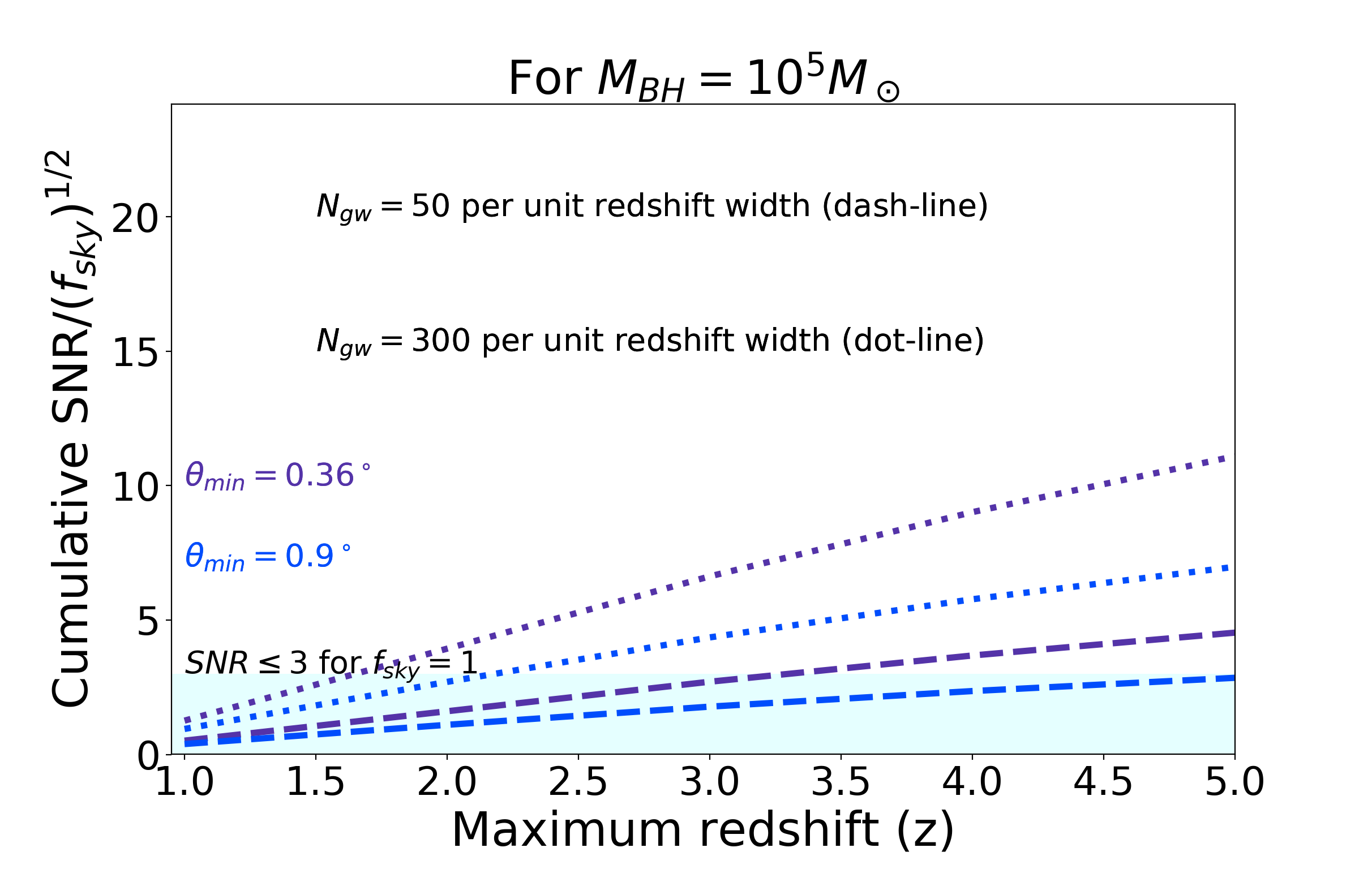}
     \caption{}\label{Fig-cmb-gw-snr}
     \includegraphics[trim={0 0 0 0cm}, clip, width=1.\linewidth]{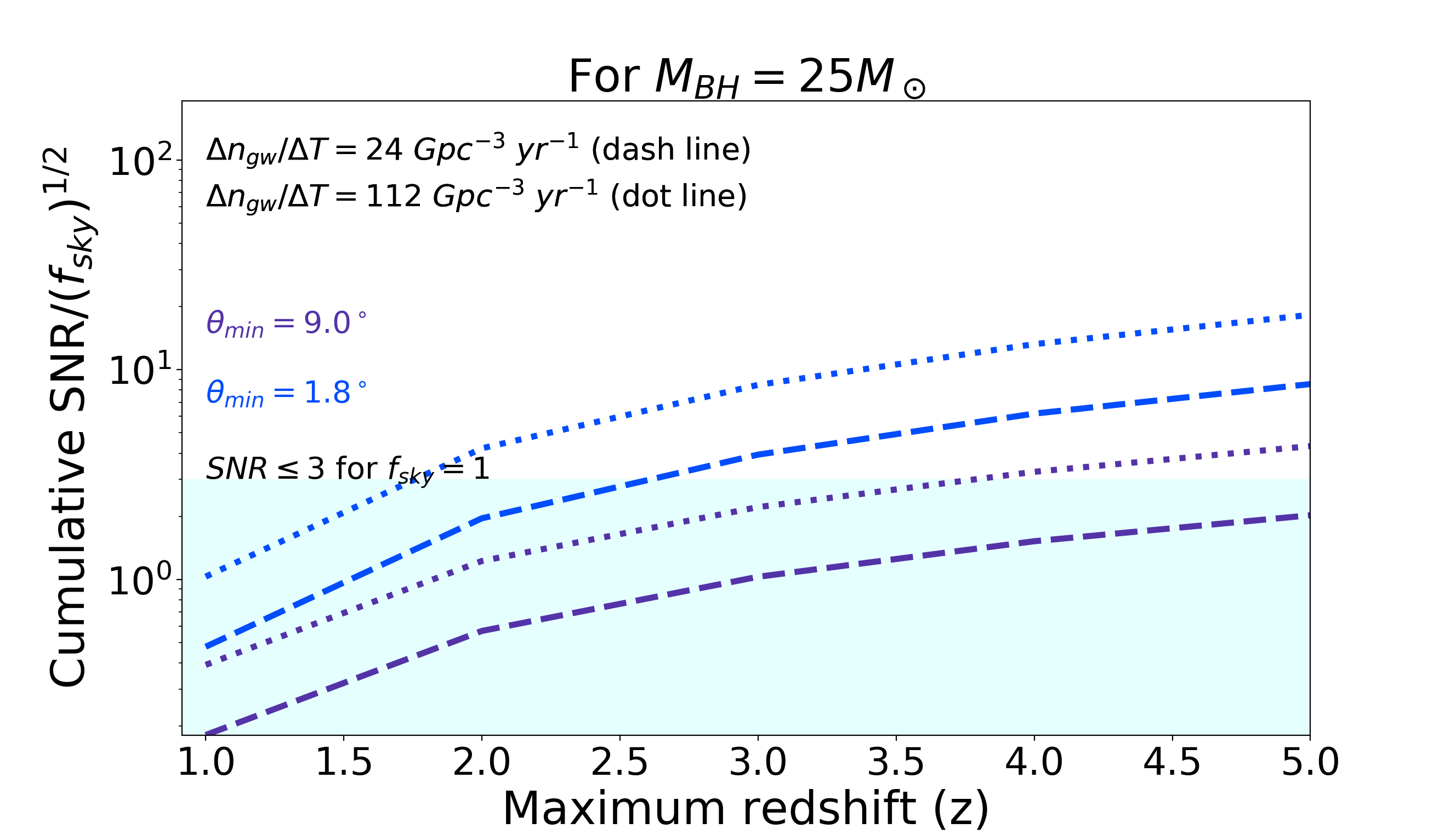}
     \caption{}\label{Fig-cmb-gw-snr-ground}
     \end{subfigure}
 \captionsetup{singlelinecheck=on,justification=raggedright}
 \caption{We show the cumulative $\text{SNR}$ as a function of the maximum cosmological redshift (z). We have considered only the inspiral phase of the unit mass-ratio BBH with total masses (a) $2\times 10^5 M_\odot$ and (b) $50\,M_\odot$. The area within the shaded region has SNR less than three for $f_{sky}=1$.} \label{Fig-cmb-gw-snr}
\end{figure}
\paragraph{Conclusion :}
The cross-correlation of CMB-photons and GW signal can be a path-breaking probe of fundamental physics. First, to observe the expected correlation proves that the GWs and EWs propagate on identical spacetime geodesics. The existence of a non-zero correlation between these two signals will manifestly verify a fundamental prediction of general relativity. Secondly, the strength of the correlated signal will also probe alternative theories to general relativity. Under the framework of general relativity and the LCDM model of cosmology, the predicted signal as a function of GW source redshift is shown in Fig. \ref{fig-cmbgwsc}. Any variation from this predicted signal for the known BBHs distributions will be a signature of alternative theories of general relativity \cite{Saltas:2014dha, Nishizawa:2017nef, Cardoso:2002pa}. Thirdly, this probes the gravitational influence of matter on GWs, or graviton-graviton interactions in the perturbation regime \cite{Delfino:2012aj}.  {Our approach also makes it possible to measure any deviation in the scalar potentials $\Phi$ and $\Psi$ from the prediction of general relativity.} This method can also be more generally applied to study the cross-correlation of the GW signal from neutron star binaries, black hole-neutron star binaries and with other probes of cosmic density field such as galaxy surveys \cite{Mukherjee:2019galgw}.

\textbf{Acknowledgement}
The authors would like to acknowledge the use of the LISA sensitivity curve tool \cite{lisa-noise}. S.M. would like to thank Karim Benabed, Luc Blanchet, Neal Dalal, Irina Dvorkin, Zoltan Haiman, David Spergel and Samaya Nissanke for useful inputs. B.D.W. would like to thank Tessa Baker for insightful discussions on alternative theories of gravity and pointing to an important references. S.M. and B.D.W acknowledge the support of the Simons Foundation and the Labex ILP (reference ANR-10-LABX-63) part of the Idex SUPER, and received financial state aid managed by the Agence Nationale de la Recherche, as part of the programme Investissements d'Avenir under the reference ANR-11-IDEX-0004-02. BDW thanks the CCPP at NYU for hospitality during the completion of this work. We have used the  following packages in this analysis: CLASS \cite{2011arXiv1104.2932L, 2011JCAP...10..037A, 2011JCAP...07..034B}, IPython \cite{PER-GRA:2007}, Mathematica \cite{mathematica}, Matplotlib \cite{Hunter:2007},  NumPy \cite{2011CSE....13b..22V}, and SciPy \cite{scipy}.

\bibliography{CMB-GW-prl}
\bibliographystyle{apsrev}
\end{document}